# Giant Seebeck coefficient in semiconducting single-wall carbon nanotube film

Yusuke Nakai[1*], Kazuya Honda[1], Kazuhiro Yanagi[1], Hiromichi Kataura[2],

Teppei Kato[3], Takahiro Yamamoto[3,4], and Yutaka Maniwa[1*]

[1] *Department of Physics, Faculty of Science and Engineering, Tokyo Metropolitan University, Hachioji, Tokyo 192-0397, Japan*

[2] *Nanosystem Research Institute (NRI), National Institute of Advanced Industrial Science and Technology (AIST), Tsukuba 305-8562, Japan*

[3] *Department of Electrical Engineering, Graduate School of Engineering, Tokyo University of Science, Katsushika, Tokyo 125-8585, Japan*

[4] *Department of Liberal Arts (Physics), Faculty of Engineering, Tokyo University of Science, Katsushika, Tokyo 125-8585, Japan*

E-mail: nakai@tmu.ac.jp, maniwa@phys.se.tmu.ac.jp

**Abstract**

We found a giant Seebeck effect in semiconducting single-wall carbon nanotube (SWCNT) films, which exhibited a performance comparable to that of commercial $Bi_2Te_3$ alloys. Carrier doping of semiconducting SWCNT films further improved the thermoelectric performance. These results were reproduced well by first-principles transport simulations based on a simple SWCNT junction model. These findings suggest strategies that pave the way for emerging printed, all-carbon, flexible thermoelectric devices.

Most widely used commercial thermoelectric materials such as $Bi_2Te_3$ and its alloys have $ZT = S^2T/\rho\kappa \sim 1$, where $S$, $\rho$, $\kappa$ and $T$ are the Seebeck coefficient, electrical resistivity, thermal conductivity and absolute temperature, respectively. These inorganic high-$ZT$ materials, however, have intrinsic disadvantages, such as the use of low-abundance and/or heavy elements, the difficulty of processing because of their brittleness, and toxicity issues. Recent studies show that the conducting polymer poly(3,4-ethylenedioxythiophene):poly(styrenesulfonate) exhibits relatively high $ZT$ values of 0.2 − 0.4.[1] These polymers are ideal for thermoelectric materials because of their flexibility and light weight, but they suffer from corrosion issues that can be a severe disadvantage for practical use. On the other hand, single-wall carbon nanotubes (SWCNTs) possess many desirable properties from a thermoelectric engineering standpoint.[2] Despite such novel properties, the $ZT$ reported for bulk SWCNTs has remained in the range of $10^{-3}$ to $10^{-4}$. This is due mainly to their high thermal conductivity and low Seebeck coefficient.[3–9] For example, an $S$ of up to only about 60 μV/K was reported for SWCNT films with an uncontrolled mixture of semiconducting and metallic SWCNTs. An enhanced Seebeck coefficient of 260 μV/K was observed only in a junction between an individual SWCNT and a metal electrode.[10] Recently, a technique for large-scale preparation of highly-enriched semiconducting and metallic SWCNTs was developed.[11, 12] It allowed us to investigate the Seebeck coefficient for the first time. Here we report the results on five SWCNT films with different semiconducting ratios, combined with first principles transport simulations of their thermoelectric properties. These results open up new possibilities, such as printed all-carbon thermoelectric devices that are flexible, lightweight, chemically stable and mechanically strong.

Five sheets of SWCNT films with different metallic-semiconducting ratios were prepared. The initial SWCNTs (Arc-SO,), which are mixtures of semiconducting and metallic SWCNTs, were purchased from Meijo-Nanocarbon Co. and dispersed into 2 weight % deoxycholate sodium salt solutions. SWCNTs with a controlled semiconducting ratio were obtained through density gradient ultracentrifugation.[12] Optical absorption spectra were used to determine the semiconducting fraction $\alpha$.[13] The films were also characterized by powder X-ray diffraction measurements using synchrotron radiation X-rays with a wavelength of 0.100 nm at BL-8A and 8B in PF, KEK, Raman spectroscopy, and scanning probe microscopy (Shimadzu SPM-9600). These measurements indicate that the mean SWCNT diameter is 1.44 nm and that many SWCNTs condense into bundles with a thickness of 5-30 nm in diameter. The film thickness is in the range of 50-130 μm. The filling factor of the SWCNTs was estimated as 0.5-0.8 from the weight and dimensions of the films. Standard four-probe measurements of the electrical resistivity and Seebeck coefficient were conducted using a physical property measurement system from Quantum Design and also using our homemade setup. The resistivity was calculated from the film dimensions and measured resistance, without correction for the filling factor of SWCNT bundles.

Figure 1(a) displays the temperature dependence of the electrical resistivity of semiconducting, mixed, and metallic SWCNT films. The resistivity of metallic SWCNTs is almost constant with temperature, whereas that of semiconducting SWCNTs is one order of magnitude larger than that of metallic ones and exhibits divergent behavior at low temperatures, which are consistent with a previous report.[13] The $T$ dependence of $S$ is shown in Fig. 1(b). The sign of $S$ is always positive, indicating hole-like carriers. The $S$ of the semiconducting SWCNT films increases upon heating and does not show the $1/T$-like dependence typical of non-degenerate semiconductors. It also monotonically increases with increasing $\alpha$, reaching approximately 170 μV/K at $\alpha \sim 1$. This is comparable to that of optimally doped $Bi_2Te_3$ and is the highest among carbon nanotube films or mats, as far as we know. In contrast, the small $S$ and the $T$-linear-like dependence of the metallic SWCNT film are typical of usual metals. The $S$ value of mixed SWCNT film is between those for $\alpha \sim 0$ and 1, and is consistent with those reported for SWCNT samples with a nominal fraction $\alpha = 67\ \%$.[3]

Figure 2(a) summarizes the $\alpha$ dependence of $S$ at 300 K. The power factor $P = S^2/\rho$ at 300 K is 33, 6.7 and 3.9 µW/mK$^2$ for semiconducting, mixed, and metallic SWCNT films, respectively. Because $S$ increases and $\rho$ decreases upon heating, a better $P$ would be obtained at higher temperatures.

The large difference in the Seebeck coefficient $S$ between semiconducting and metallic SWCNTs is naturally ascribed to their electronic structures. It is known that as-prepared SWCNT films have already been doped with $p$-type carriers; one of the most probable doping is caused by $O_2$ physisorption during sample preparation and/or exposure to air. It shifts the Fermi level $E_F$ and changes the density of states (DOS) at $E_F$. In semiconducting SWCNTs, this change is much more significant than in metallic SWCNTs, because $E_F$ moves toward the Van Hove singularity point of the DOS in semiconducting SWCNTs with little carrier doping. Thus, semiconducting SWCNTs exhibit a larger magnitude of $S$ than metallic SWCNTs.

The observed dependence of $S$ on the fraction $\alpha$ suggests that all-SWCNT thermoelectric devices can be fabricated simply. We made a SWCNT thermoelectric device as shown in Fig. 3, which basically consists of two types of SWCNT films: SWCNT films with $\alpha > 90\ \%$ and those with the as-prepared concentration of $\alpha \sim 67\ \%$. A temperature difference of approximately 10 K generates 2.6 mV. Because the device consists of a series of 10 pairs, each pair generates 0.26mV, corresponding to 26 µV/K for a pair. Although the performance is about one-fourth of the expected *S*, it should be improved by optimization of the film fabrication process. Such a device is basically fabricated using a printing technique.[2]

In addition to a high *S*, ideal thermoelectric materials need a low thermal conductivity $\kappa$. Therefore, the exceptionally high $\kappa$ of individual SWCNTs may become troublesome.[14,15] The high $\kappa$ of individual SWCNTs, however, is not directly reflected in the observed $\kappa$ in bulk SWCNT films, which depends significantly on the morphology of the bulk SWCNTs, such as the degree of SWCNT alignment and SWCNT length. The reported values ranges from 0.1 to 250 W/mK in bulk SWCNTs.[4, 16] These values, even taking into account the volume filling of SWCNTs in films, are much smaller than the intrinsic $\kappa$ of SWCNTs which may be larger than 2000 W/mK.[14, 17] Thus the observed $\kappa$ is believed to be limited by highly resistive thermal junctions present in films. For example, these junctions can be formed not only within bundles but also between bundles (see Fig. 2(b)).[16]

The observed giant Seebeck effect cannot be ascribed solely to the intrinsic properties of SWCNTs because a temperature variation in bulk SWCNTs almost always occurs at the thermally resistive junctions. Considering a difference of 10 times between the intrinsic and observed $\kappa$, naively the *S* of each SWCNT must be 10 times larger than the observed *S* if it is ascribed to the intrinsic nature of SWCNTs. This seems to be improbably large, so an alternative mechanism is required. We propose below a different mechanism by which the observed giant Seebeck effect arises from SWCNT junctions.

We now estimate the value of *ZT*. By using a reported value of $\kappa$ = 0.15 W/mK for a bulk SWCNT sample with a similar structure to our sample (non-aligned SWCNTs with a mean diameter of 1.44 nm),[16] we obtain a *ZT* of 0.13 at 350 K. Assuming that the thermal conductivity is proportional to the SWCNT length,[18] *ZT* can be much larger than 0.33 because the SWCNTs used in the present study were 0.5-2 μm in length, whereas those in Ref. 16 were 5-15 μm. These *ZT* values are comparable to or larger than those of the best organic thermoelectric materials.[1]

Furthermore, to tune the carrier number, we performed high-temperature annealing under dynamic vacuum for de-doping and acid treatments for hole-doping in the semiconducting films with $\alpha \sim 1$. Because as-grown films are presumably doped with $O_2$,[5] high-temperature annealing was performed for de-doping under dynamic vacuum of $10^{-4}$ Pa. For *p*-type doping, the semiconducting films were acid-treated as follows. First, we annealed the semiconducting SWCNT films at 500 °C for 20 min under dynamic vacuum to remove adsorbed molecules. Then the films were effectively *p*-doped via chemical treatment with HCl, $HNO_3$, $H_2SO_4$,[19, 20] through either physisorption on the SWCNT surface or chemisorption with hydroxyl (-OH) or carboxyl (-COOH) formation on dangling bonds or defects.[21] The films were immersed in 11.7 M HCl, 13.1 M $HNO_3$, 18.1 M $H_2SO_4$ for 1 h respectively, and dried overnight in the atmosphere. For the $HNO_3$-doped SWCNT films, pumping at 30-100 °C was also used to control the hole-density in a step by step manner. Figure 4(a) shows that both the high-temperature annealing and acid treatment substantially reduced *S*. In contrast, $\rho$ increased upon de-doping and decreased on hole-doping via the acid treatments, as reported in SWCNT mats with $\alpha \sim 0.67$.[8, 19] It is interesting to know the relationship between the power factor $P = S^2/\rho$ and the resistivity $\rho$ as shown in Fig. 4(b). The semiconducting SWCNT film shows a better *P* than the mixed film. Second, we find that the semiconducting SWCNT film exhibits a maximum *P* in the acid-treated region, $P = 108$ μW/mK$^2$, which corresponds to a *ZT* of 0.22 at 300 K by using $\kappa = 0.15$ W/mK. Here, $\kappa$ is assumed to be independent of carrier doping because $\kappa$ is dominated by phonon contributions.[4] This improvement in *ZT* is due mainly to the large reduction in the resistivity by doping while maintaining a moderate magnitude of *S*.

Clarifying the mechanism of the thermoelectric properties is important for designing high-performance thermoelectric materials. As already mentioned, the heat-resistive junctions in SWCNT films, i.e., intra- and inter-bundle junctions, may be responsible for the large $S$. Here, we examine the following simplified models to analyze the $S$ vs. $\alpha$ relationship displayed in Fig. 2. The first model represents a serial junction of SWCNTs, as shown in Fig. 2(b).[7, 22] There are three types of junction in the model: semiconducting-semiconducting (s-s) SWCNT junctions with the Seebeck coefficient $S_{ss}$, semiconducting-metallic (s-m) junctions with $S_{sm}$, and metallic-metallic (m-m) junctions with $S_{mm}$. Here, we use $S_{ss} \approx 12.1 S_{mm} = 170$ µV/K, which was estimated from the results at $\alpha = 0$ and 1, and the number of s-s junctions is proportional to $\alpha^2$. We also assume $S_{sm} \approx S_{mm}$ and the same temperature difference $\Delta T_J$ at all the junctions. Then the Seebeck coefficient of the serial model is given by $S_s \approx \alpha^2 S_{ss} + (1-\alpha^2) S_{mm}$. The second model represents a parallel junction where the s-s, s-m and m-m junctions are connected in parallel. In the parallel model, $S$ is given by an electrical conductance-weighted average,[7, 22]

$$S_p = \frac{1}{\sigma_{Jm}(1-\alpha^2) + \sigma_{js}\alpha^2} \left[ S_{mm}(1-\alpha^2)\sigma_{Jm} + S_{ss}\alpha^2 \sigma_{Js} \right], \tag{1}$$

where $\sigma_{Jm}$ and $\sigma_{Js}$ are the junction conductance for the m-m (or m-s) and s-s junctions, respectively. In the third model, $S$ is given by the sum of the values for the serial and parallel junctions with a fitting parameter $\beta$, $S = (1-\beta)S_s + \beta S_p$. The results are shown as lines in Fig. 2(a). The serial (thin solid line) and parallel (thick solid line) models roughly describe the observed $S$, but these calculated results do not fit well with the observed $S$. The best result is obtained from the third model with $\beta = 0.6$ and $\sigma_{Jm} = 5.1\sigma_{Js}$, suggesting that the junction model consisting of serial and parallel junctions reproduces the observed $S$ vs. $\alpha$ relationship well.

To verify that the remarkable thermoelectric performance of highly-enriched semiconducting SWCNT films is caused by tube-tube contacts, we performed first-principles transport simulations of a semiconducting SWCNT junction, as shown in Fig. 5(a). The Seebeck coefficient and electrical resistance were calculated using Landauer formula generalized for the thermoelectric properties.[22] According to the generalized Landauer formula, not only the electrical resistance (or electrical conductance) but also the Seebeck coefficient can be expressed by the transmission function. We calculated the transmission function of the SWCNT junction using Atomistix ToolKit (ATK 12.2.0),[23] which is based on density functional theory combined with the non-equilibrium Green's function method. This software employs numerical atomic orbitals and norm-conserving pseudopotentials. We used a double-zeta plus polarization basis set and the local density approximations for the exchange-correlation potential. In our simulation, the range of the chemical potential $\mu$ was set to $\mu \approx -0.55 \sim -0.22 \mathrm{eV}$, corresponding to hole doping. Here, $\mu = 0\,\mathrm{eV}$ and $\mu = -0.28\,\mathrm{eV}$ are the Fermi energy and valence-band top of the (8,0) SWCNT, respectively. Figure 5(b) displays the Seebeck coefficient $S$ (solid curve) and a power factor $\tilde{P} = S^2/R$ (dashed curve) of the junction at 300 K as a function of the resistance $R$ in units of the quantum resistance. First, we find that a fairly large $S$ can be obtained within the present junction model. Next, the filled and open circles are the experimental data in Fig. 4(b) scaled to fit the solid and dashed curves, respectively. The simulation results are in excellent agreement with the experimental data. The deviation of the experimental data from the simulation data in the high-resistance region is probably attributable to the existence of a small amount of metallic SWCNTs in the film because the observed $S$ is given by the electrical-conductance-weighted average in the parallel model.

In conclusion, we found a giant Seebeck effect in highly concentrated semiconducting

SWCNT film. The Seebeck coefficient is the largest among bulk carbon materials, as far as we know. It is suggested that *ZT* of approximately 0.33 would be further improved by optimization of the SWCNT length/diameter, bundle structures, doping level and functionalization. Furthermore, the present study clarified that the thermally resistive junctions play an important role in the giant Seebeck effect of SWCNT films. Because major advantages of SWCNTs as a thermoelectric material are their printability and flexibility, these findings represent a major advance in the realization of emerging printed flexible thermoelectric devices.

**Acknowledgements**

This work was supported in part by a Grant-in-Aid for Scientific Research on Priority Area "New Materials Science Using Regulated Nano Spaces" from the Ministry of Education, Culture, Sports, Science and Technology of Japan, and by JSPS KAKENHI Grant Numbers 21540421, 24681021, 25800201 and 25104722. This research was also supported by JST, CREST.

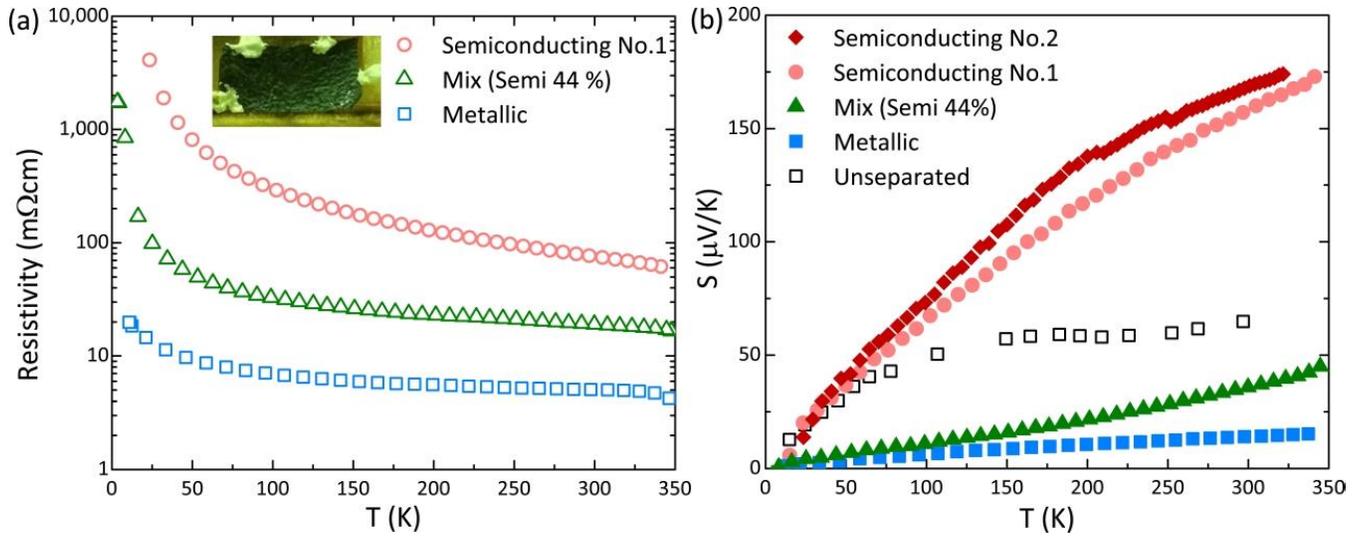

**Fig. 1.** Temperature dependence of resistivity (a) and Seebeck coefficient *S* (b) of semiconducting, mixed, and metallic SWCNT films. Fractions of semiconducting SWCNT are 100, 98, 40, and 0 % within an error of 4 % for semiconducting No.2, No.1, mixed, and metallic SWCNT films, respectively. Open squares represent as-prepared SWCNT film ($\alpha \sim 0.67$) cited from Ref. 3. Inset: Photo of SWCNT film mounted on a transport measurement holder.

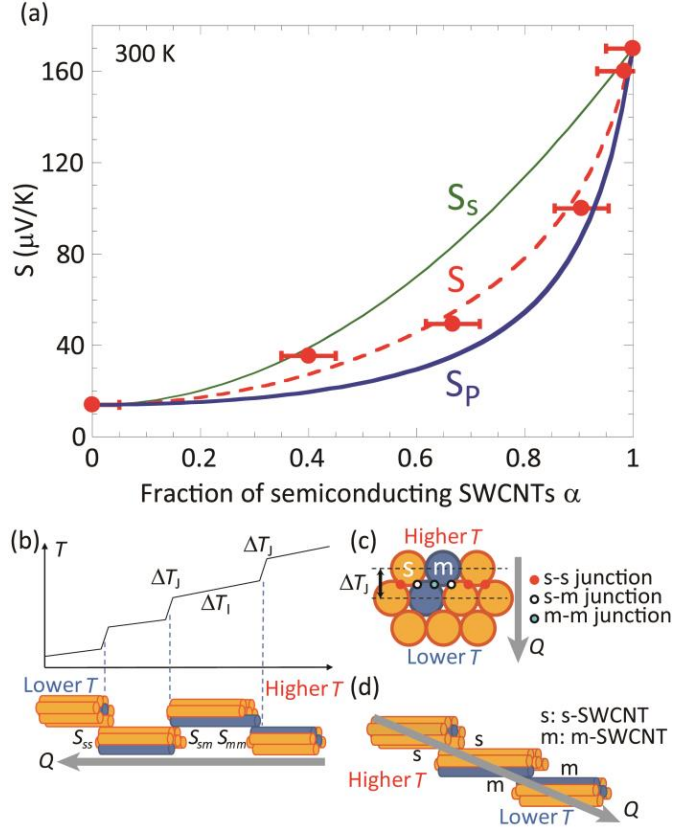

**Fig. 2.** (a) Semiconducting fraction $\alpha$ dependence of Seebeck coefficient $S$ at 300 K. Blue solid (green thin solid) line represents $S$ for parallel (serial) model with $\sigma_{Jm} \approx 5.1\sigma_{Js}$; red broken line represents their combined model with $\beta = 0.6$ (see text). (b) Schematic serial model for junction networks consisting of semiconducting (s) and metallic (m) SWCNTs, where temperature ($T$) varies along the series of SWCNT bundles. Each rod represents individual m- or s-SWCNT, which form bundle structures. There are three types of junctions, semiconducting-semiconducting (s-s), semiconducting-metallic (s-m) and metallic-metallic (m-m) junctions with Seebeck coefficients   and  , respectively. We assume that a finite temperature variation $\Delta T_J$ is induced at each junction but not at each tube ($\Delta T_I \sim 0$). (c) Parallel model, in which heat $Q$ flows perpendicular to the SWCNT tube axes. (d) Example of third model combining serial and parallel models, where $S_{ss} \approx 10\, S_{sm} \approx 10 S_{mm}$ is assumed.

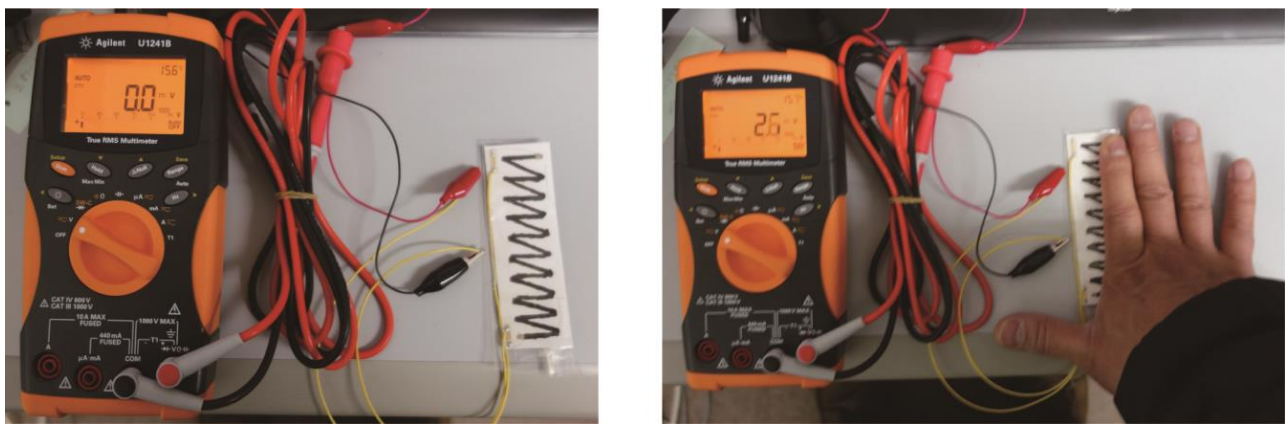

**Fig. 3.** Printed SWCNT thermoelectric device consisting of a series of 10 pairs of highly-enriched semiconducting and as-prepared SWCNT films that were mixed with carboxymethylcellulose. By placing a hand on one side of the device, 2.6 mV is generated at room temperature, as shown on the bottom.

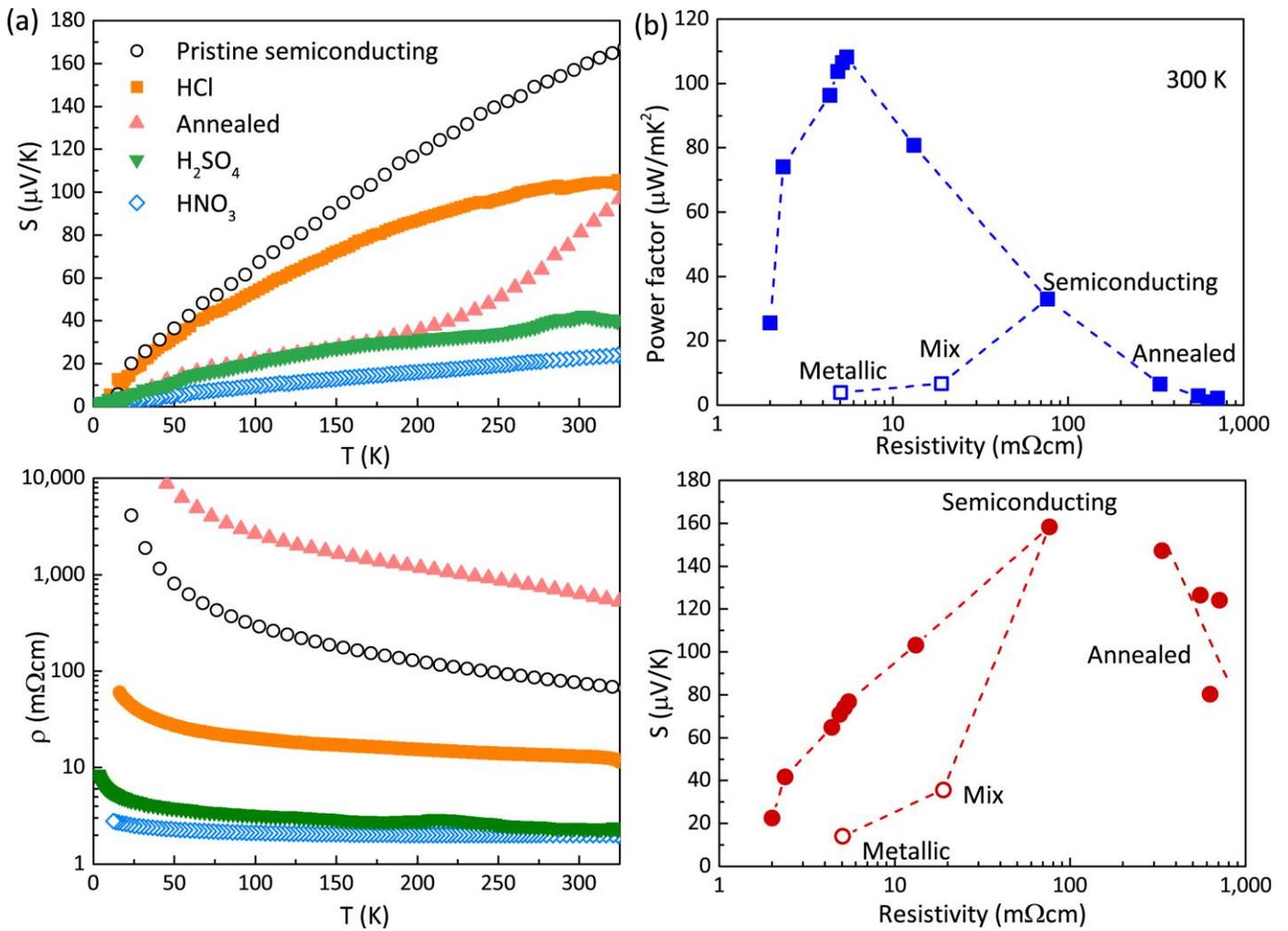

**Fig. 4.** (a) Temperature dependence of Seebeck coefficient $S$ and electrical resistivity $\rho$ and (b) power factor $P$ and $S$ as a function of $\rho$ at 300 K of acid-treated and vacuum-annealed semiconducting SWCNT films (filled symbols) and mixed and metallic SWCNT films. Lines are a guide to the eyes.

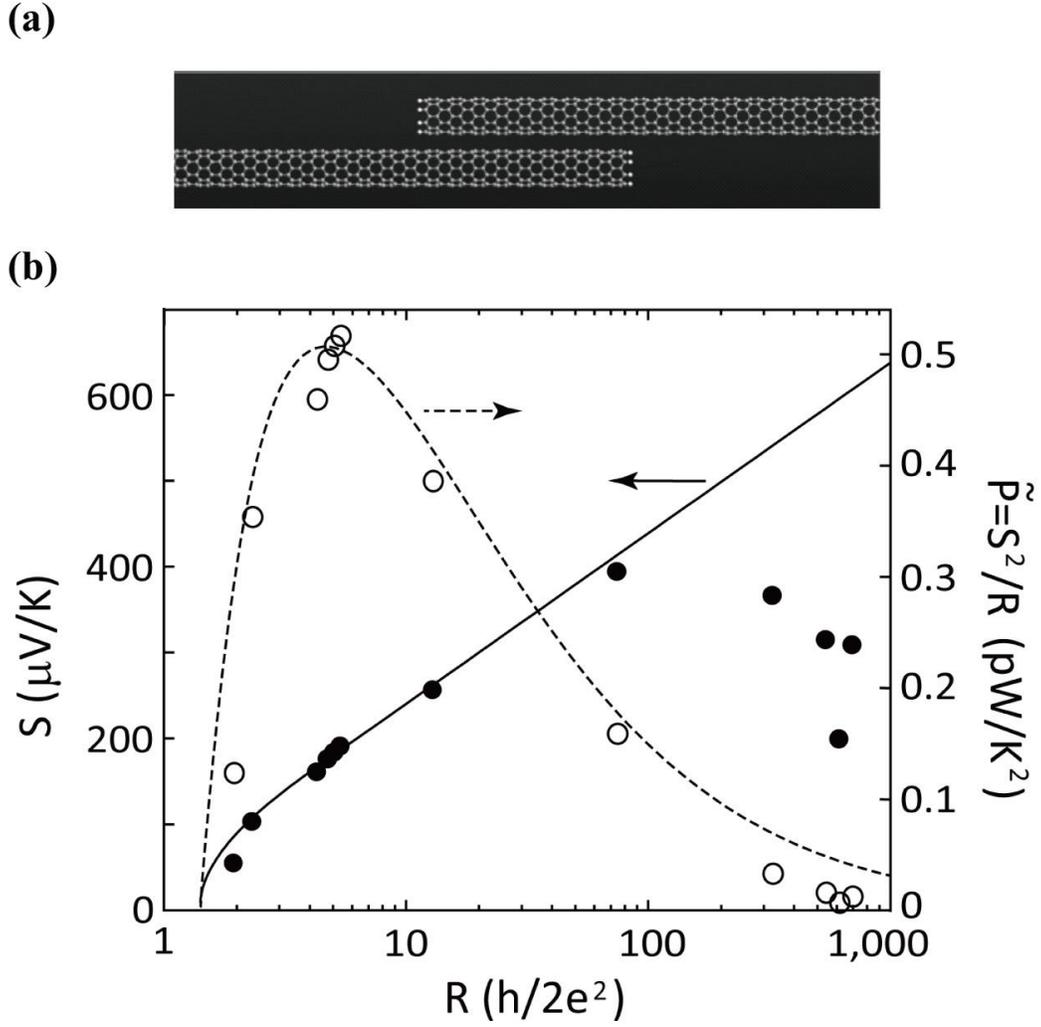

**Fig. 5.** (a) Semiconducting SWCNT junction consisting of laterally contacted (8,0) SWCNTs. The contact length and tube-tube distance were set to 38 Å and 3.35 Å, respectively. (b) Seebeck coefficient $S$ (solid curve) and power factor $\tilde{P} = S^2/R$ (dashed curve) of the (8,0)-(8,0) SWCNT junction at 300K as a function of the resistance $R$ in units of the quantum resistance $h/2e^2 = 12.9$ kΩ. Filled and open circles are experimental data for $S$ and $P$ [shown in Fig. 4(b)] scaled to fit the solid and dashed curves, respectively.